\documentclass{aa} 
\usepackage{graphicx}
\usepackage{natbib}
\bibpunct{(}{)}{;}{a}{}{,}
\defcitealias{ca89}{CCM}
\begin{document}
\title{An uncatalogued optical \ion{H}{ii} region at the outskirts of the Galaxy\thanks{Based on observations 
made with the Nordic Optical Telescope, operated on the island of La Palma jointly by Denmark, Finland,
Iceland, Norway, and Sweden, in the Spanish Observatorio del Roque de los Muchachos of the Instituto de
Astrofisica de Canarias.}}
\author{S. Temporin\inst{1}\thanks{Guest User, Canadian Astronomy Data Centre, 
which is operated by the Dominion Astrophysical Observatory for the National 
Research Council of Canada's Herzberg Institute of Astrophysics.}
\and
R. Weinberger\inst{1}
}
\offprints{giovanna.temporin@uibk.ac.at}
\institute{Institut f\"ur Astrophysik, Universit\"at Innsbruck,\\
              Technikerstra\ss e 25, A-6020 Innsbruck, Austria\\
              \email{giovanna.temporin@uibk.ac.at}\\
	      \email{ronald.weinberger@uibk.ac.at}
	     }
	     
\date{Received ?? / Accepted ??}

\abstract{We present NOT optical observations of a clump
(l = 127\fdg9435 , b = $+$1\fdg8298), embedded in an extended, irregularly
shaped, diffuse optical nebula. 
This condensation shows an emission-line spectrum typical of classic \ion{H}{ii} regions.
Although its location on the sky coincides with a nearby extended photoionized region recently 
identified by \citet{ci03} in radio data from the Canadian Galactic Plane Survey (CGPS), the systemic
velocity of this $\approx$ 1\arcmin-sized \ion{H}{ii} region, V$_{\rm {LSR}}$ = $-$71$\pm$12 km s$^{-1}$,
poses it far out in the Galaxy, beyond the Perseus arm. The location of this region in 
the Galaxy is supported by \ion{H}{i} structures visible at comparable radial velocity on
CGPS data.
We argue that this \ion{H}{ii} region might belong to an outer Galactic arm. 
The emission line ratios of the surrounding extended nebula, whose radial velocity is
consistent with that of the small \ion{H}{ii} region, are typical of photoionized gas in the
low density limit. Smaller clumps of comparable surface brightness are visible within the optical 
boundaries of the extended, faint nebula. After comparison of the optical data with far infrared 
and radio observations, we conclude that this nebula is an \ion{H}{ii} region, $\sim$ 70 pc in size, 
probably photoionized by an association of OB stars and
surrounded by a ring of neutral hydrogen.  

\keywords{(ISM:) \ion{H}{ii} regions; ISM: individual objects: IRAS 01342+6358, IRAS 01330+6351}
}
\titlerunning{Distant Galactic \ion{H}{ii} region}
\maketitle

\section{Introduction}

Galactic \ion{H}{ii} regions offer the opportunity of investigating and understanding the details of star formation
processes. Furthermore, the identification of \ion{H}{ii} regions is of interest to trace the spiral
structure of our Galaxy, which is still a matter of debate \citep[e.g.][ and references therein]{mcg04,ru03}.

During our searches for galaxies beyond the Galactic plane -- based on visual inspection of 
Palomar Observatory Sky Survey plates of both first and second generation
\citep[e.g., ][]{ssw96,wgz99} -- 
we have identified an extended and irregularly shaped nebula, not previously catalogued, within 
(or in projection onto) which a number of brighter condensations are visible. 
The whole object is extended over about half a degree on the sky, while 
the bright condensations are much more limited in size, the brightest and largest one, 
located at l = 127\fdg9435 , b = 1\fdg8298, having a major axis of $\approx$ 1\arcmin.
Its morphology happens to be similar to that of the giant ($\sim$ 2\degr$\times$2\degr\ in extent) 
\ion{H}{ii} region \object{IC 1805} (also known as W4), ionized by the association \object{Cas OB6} 
and located at a distance of 2.35 kpc \citep{mjd95}.

Two IRAS peaks of emission, listed in the IRAS Point Source Catalogue
as \object{IRAS 01342+6358} and \object{IRAS 01330+6351}, are present in the same region of the sky 
and were identified and measured as actually extended FIR sources by \citet{ft96}.
Observations of one of these sources, 
\object{IRAS 01342+6358}, during the CS(2-1) survey of ultracompact (UC) \ion{H}{ii} regions \citep{bnm96} left the source 
undetected, therefore its radial velocity and
kinematic distance could not be determined. Recently, \citet{ci03} gave evidence that the extended
FIR emission is associated with an extended thermal source of radio emission. They argued that 
this structure is physically related to a ``hole'' in the \ion{H}{i} distribution observed at a 
systemic velocity of $\sim$ $-$29.5 km s$^{-1}$, and traces gas photoionized by the nearby O star \object{HD
10125}. However, they do not provide any direct measurement of the radial velocity (and therefore distance)
of the ionized material. To our knowledge, no spectroscopic observations of this ionized region have been
published to date.
In this paper we present for the first time $R$-band photometry and optical spectroscopy of a portion 
of the optical extended nebula spatially coincident with the extended FIR and radio emission, 
including the $\sim$ 1\arcmin-size brighter condensation cited above.
Beside investigating the physical properties of this ``condensation'', that is revealed as a classic
\ion{H}{ii} region, we derive its radial velocity and try to establish whether this region is 
related to the surrounding extended nebula and associated to the extended photoionized
region recently identified by \citet{ci03}. 

\section{Observations and measurements}

The optical data were obtained in 1999 November at the Nordic Optical Telescope (NOT) with the faint object
spectrograph and camera ALFOSC and consist of a 330 s $R$-band exposure covering a field of view of
$\sim$ 5\arcmin$\times$5\arcmin\ and a 1800 s long-slit
spectrum taken along the major axis of the bright optical condensation (position angle = 165\degr).
Both the image and the spectrum have a spatial scale of 0\farcs188\ pixel$^{-1}$. The seeing
during the observations was $\sim$ 0\farcs8. The long-slit spectrum was taken with a
1\arcsec-wide slit and has a dispersion of 1.483 \AA\ pixel$^{-1}$ and a spectral resolution
of $\sim$ 6.5 \AA\ over the wavelength range $\lambda$ 3833 -- 6857 \AA, as evaluated from the FWHM of 
comparison lines. 
$R$-band images of a \citet{la92} standard field for photometric calibration and
long-slit spectra of the spectrophotometric standard star BD+284211 were taken three and two times,
respectively, during the same night. Standard reduction steps were performed with the available
IRAF\footnote{IRAF is distributed by the National Optical Astronomy Observatories, which are operated 
by the Association of Universities for Research in Astronomy, Inc., under cooperative agreement with the 
National Science Foundation.} packages. The 2$\sigma$ limiting magnitude and surface brightness of our $R$-band image are 
$R$ = 23.88 mag, $\mu_R$ = 24.74 mag arcsec$^{-2}$.

\subsection{Photometry}

\begin{figure*}
 \centering
 \includegraphics[width=\textwidth]{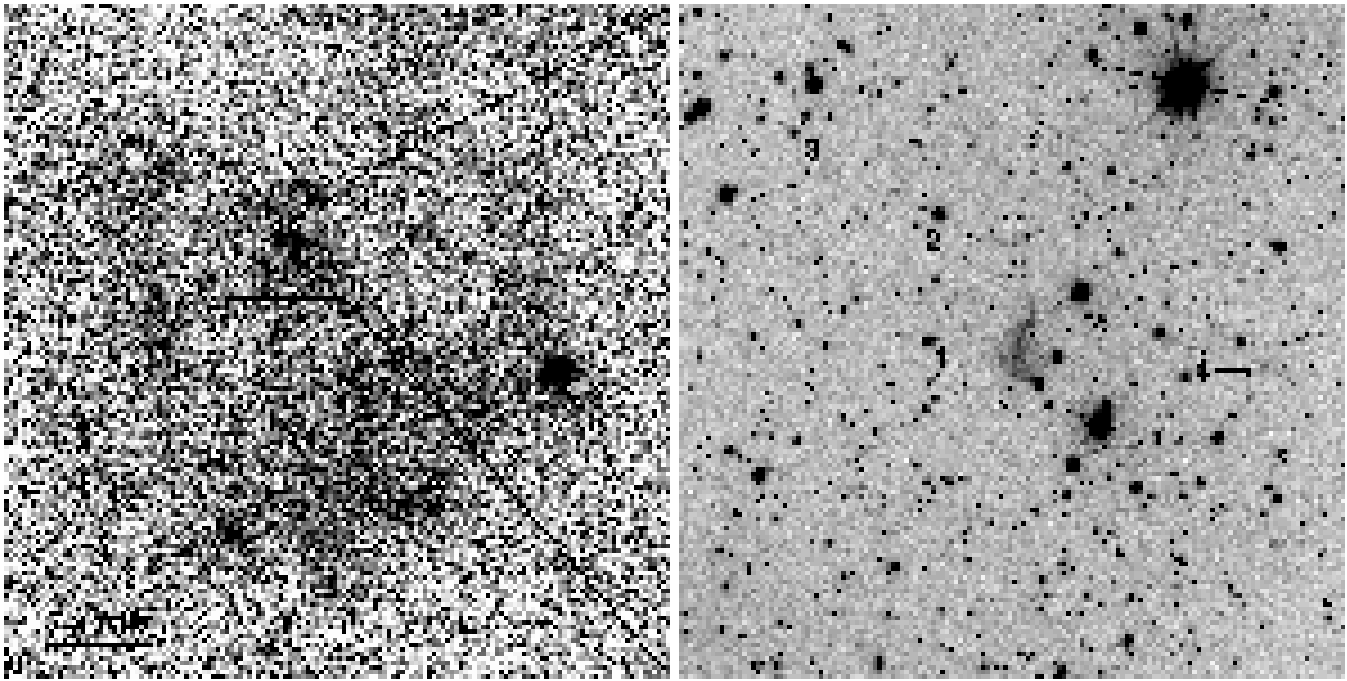} 
 \caption{\emph{Left:} R band DSS2 image of the whole nebula. North is up and East to the left. 
 \emph{Right:} Zoom on a $\sim$ 5\farcm4$\times$5\farcm4 section of the nebula, centered on the
 1\arcmin-sized bright condensation (labelled 1). Lower surface brightness small clumps are marked in the 
 northern and western sides. The image has been obtained at the NOT in the $R$ filter.}\label{dssR}
 \end{figure*}

Since the diffuse nebula, as visible in the Digitized Sky Survey image (Fig.~\ref{dssR}), extends over
about half a degree on the sky, we could image only a portion of it, which completely filled the ALFOSC
field-of-view. This fact prevented us from any estimation of the real sky-background, as well as an
evaluation of the average surface brightness of the most diffuse portion of the nebula.
Therefore, we only were able to evaluate the magnitude and surface brightness of the bright
condensations falling within our field-of-view.
A bidimensional polynomial fit to the sky-background (thus including the contribution from 
the more diffuse part of the nebula) was obtained with MIDAS\footnote{MIDAS is developed and maintained by
the European Southern Observatory.} on the R-band image after 
masking (IRAF task IMEDIT) of badly saturated stars in the field and 
subtraction of the unsaturated stars through a fit to a model point spread function (obtained with
the DAOPHOT package within IRAF). These operations were necessary because of the relatively strong 
contamination by stars, owe to the low Galactic latitude of the target. Photometric measurements 
of the clumps within the nebula  were performed on the background-subtracted, free of stars, 
image by use of both circular apertures completely encompassing the visible clumps and polygonal 
apertures following the outermost visible contours. Both methods gave similar values of total magnitudes.
The polygonal apertures were used to estimate the surface brightness. The measured values
are given in Table~\ref{phot}. 
The whole structure, except for a filamentary extension to the north, fits into $\approx$ 2/3 of a 
circle $\sim$ 15\farcs8 in radius. In other words, the bulk of
 the structure is confined within a region $\sim$ 31\farcs6 $\times$ 25\farcs6 in size.
 When the northern extension is considered as well, the major axis of the structure reaches $\sim$
 49\arcsec.
 
 \begin{table}
      \caption[]{Photometric measurements of the clumps}
         \label{phot}
 $$
   \begin{array}{lllll}
            \hline
	    \hline
            \noalign{\smallskip}
           Id. & \alpha \,(\mathrm J2000) \, \, & \delta \,(\mathrm J2000) \, \,& R  &  \mu_R \\
            & ({^\mathrm h}\,\, {^\mathrm m}\, \,{^\mathrm s})& (\degr \, \,\arcmin \,\, \arcsec)&  (\mathrm{ mag}) \,& (\mathrm{
	    mag\, arcsec}^{-2})\\
	    \noalign{\smallskip}
            \hline
            \noalign{\smallskip}
1&  01\, 37\, 48.2 & 64\, 14\, 45 & 15.53\pm0.01 &  23.23  \\
2&  01\, 37\, 49.7 & 64\, 15\, 39 & 19.90\pm0.05 &  24.51  \\
3&  01\, 37\, 58.2 & 64\, 16\, 24 & 20.04\pm0.04 &  24.28  \\
4&  01\, 37\, 29.1 & 64\, 14\, 31 & 18.72\pm0.02 &  24.62  \\
            \noalign{\smallskip}
            \hline
         \end{array}
     $$ 
   \end{table}

\subsection{Spectroscopy: Physical properties of the region}

\begin{figure}
\centering
\includegraphics[width=\linewidth]{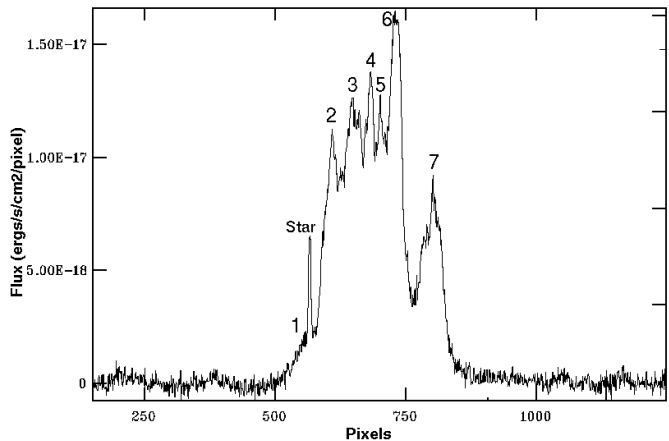} 
\caption{H$\alpha$ emission-line profile along the slit, after subtraction of the sky-background. Individual
subregions and a peak due to an overlapping star are marked.}  \label{prof}
\end{figure}

The long-slit spectrum showed extended line emission without any detectable continuum. 
Considerable substructure was detected in the source,
as it can be seen from the H$\alpha$ emission-line profile along the slit (after background subtraction) 
in Fig.~\ref{prof}, where individual subregions are numbered 1 to 7. 
We measured emission line fluxes both in the
individual subregions and in the total spectrum of the source in order to determine possible 
variations of physical properties across it, along with its average properties.
Both the observed and extinction corrected emission line fluxes are reported in Table~\ref{fluxes},
together with the amount of foreground extinction as derived from the Balmer decrement assuming as
an intrinsic ratio H$\alpha$/H$\beta$ = 2.86, the theoretical value for case B recombination nebulae
\citep{ost89}. The total foreground extinction to the source, expressed in terms of color excess E($B-V$),
ranges between $\sim$ 0.5 and 1.0 mag. The foreground extinction caused by intervening Galactic material
along the line-of-sight to the source is difficult to determine. However, considering the 12 stars 
with known spectral type, estimated distances in the range 1.17 to 4.28 kpc, and projected distances 
within a radius of 1\degr\ from the source, we found an
average extinction E($B-V$) = 0.5$\pm$0.3 mag, in rough agreement with the absorption of 1.0$\pm$0.2 mag
derived by \citet{ci03} considering the \ion{H}{i} column density out to velocities around 
$-$30 km s$^{-1}$. Therefore we estimate for the observed source an internal extinction 
E($B-V$) $\la$ 0.5 mag.

\begin{figure}
\centering
\includegraphics[width=\linewidth]{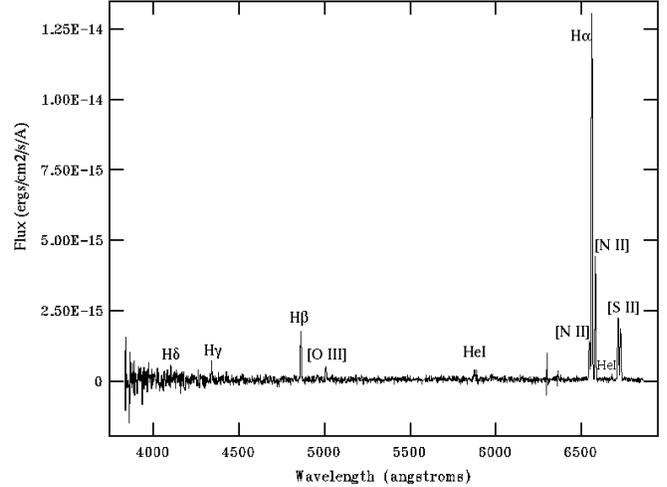} 
\caption{Total spectrum of the \ion{H}{ii} region. All detected emission lines
are marked.}  \label{spec_tot}
\end{figure}

The classic diagnostic emission-line ratios (Table~\ref{ratios}) we measured are all typical of 
photoionized \ion{H}{ii} regions \citep[e.g.][ and references therein]{gwb99}. No evidence of ionization 
by shocks was found.
Unfortunately the lack of a detectable [\ion{O}{iii}] $\lambda$4363 line prevented us from
estimating the electronic temperature of the ionized gas, therefore we assumed a canonical
value for \ion{H}{ii} regions, T$_{\mathrm e}$ = 10$^4$ K and derived the electronic density 
N$_{\mathrm e}$ from the ratio of the [\ion{S}{ii}] $\lambda\lambda$6716,6731 lines.
We found N$_{\mathrm e}$ values in the range 1.7 - 2.4$\times$10$^2$ cm$^{-3}$ (Table~\ref{fluxes}), which
are characteristic of classic \ion{H}{ii} regions and at least one to two orders of magnitude lower 
than values typical of compact or UC\ion{H}{ii} regions \citep[e.g.][ and references therein]{fkg03}.
The total spectrum of the \ion{H}{ii} region is shown in Fig.~\ref{spec_tot}. 
The presence of very strong [\ion{O}{i}] $\lambda\lambda$6300,6364 
night-sky lines prevented us from establishing with certainty the presence or absence of [\ion{O}{i}] 
emission from the nebula. However, a comparison of the [\ion{O}{i}] lines on- and off-source along the slit 
seems to indicate that these emission lines are absent or extremely week on the spectrum of the source.
Therefore, we conclude that oxygen is almost fully ionized in this region indicating that the partially
ionized zone of the nebula, where \ion{O}{i} is present, is probably a very thin layer, as expected in case
of ionizing stars of OB spectral types.

 \begin{table*}
   \scriptsize   
   \centering 
   \caption[]{Emission line fluxes$^{\mathrm{a}}$}
         \label{fluxes}
 $$
   \begin{tabular}{lllllllll}
            \hline
	    \hline
            \noalign{\smallskip}
 Line  & Part 1 & Part 2 & Part 3 & Part 4 & Part 5 & Part 6 & Part 7 & Total\\
	    \noalign{\smallskip}
            \hline
            \noalign{\smallskip}
H$\delta$ & ...& 2.922$\pm$0.463& 3.903$\pm$0.333& 3.459$\pm$0.497& 1.358$\pm$0.278& 3.488$\pm$0.257&... &
15.840$\pm$0.164\\
          & ...& 206.50& 206.06& 193.53& 69.892& 136.41& ...& 822.19\\  
H$\gamma$ & ...& 4.933$\pm$0.202&3.743$\pm$0.312 &4.673$\pm$0.274 & 3.528$\pm$0.177& 7.570$\pm$0.316&
6.072$\pm$0.262& 28.960$\pm$0.201\\
          & ...& 270.72& 156.14& 205.87& 143.68& 238.11& 287.98& 1188.9\\
H$\beta$  & 6.325$\pm$0.190& 22.530$\pm$0.055&23.120$\pm$0.041 & 15.270$\pm$0.043&13.700$\pm$0.044 &
30.480$\pm$0.037& 22.870$\pm$0.097& 135.50$\pm$0.039\\
          & 136.01& 719.06& 582.09&403.03 & 337.85&601.20 &643.38 & 3364.7\\
$[$\ion{O}{iii}$]\, \lambda4959$ & 4.569$\pm$0.473& 5.425$\pm$0.351& 2.441$\pm$0.467&2.340$\pm$0.282 &
1.836$\pm$0.423&1.958$\pm$0.454 &... & 22.020$\pm$0.358\\
          & 90.811& 158.42& 56.575& 56.787&41.703 & 35.776& ...& 503.54\\
$[$\ion{O}{iii}$]\, \lambda5007$ & 7.446$\pm$0.205& 8.199$\pm$0.171& 5.730$\pm$0.165 &5.455$\pm$0.131
&3.335$\pm$0.157 & 5.770$\pm$0.238& ...& 40.430$\pm$0.150\\
          & 142.62& 229.65& 127.74& 127.26&72.882 & 101.71&... &889.43 \\
\ion{He}{i} $\lambda5876$ & 3.015$\pm$0.338& 4.423$\pm$0.253& 4.077$\pm$0.264&
3.211$\pm$0.232&2.411$\pm$0.261 &7.979$\pm$0.238 &... & 33.030$\pm$0.214\\
          & 34.766& 69.865& 53.309& 43.595& 31.009& 85.889& ...& 427.16\\
$[$\ion{N}{ii}$]\, \lambda6548$ &2.330$\pm$0.371 &18.380$\pm$0.044 & 19.370$\pm$0.031& 12.700$\pm$0.033&
12.660$\pm$0.036& 25.720$\pm$0.029&21.680$\pm$0.075 & 112.50$\pm$0.032\\
          & 20.229& 210.75& 187.93& 127.38& 121.05& 210.12&227.45 & 1080.9\\
H$\alpha$ & 45.080$\pm$0.019& 180.70$\pm$0.004& 172.70$\pm$0.004& 115.70$\pm$0.004&101.70$\pm$0.004
&211.70$\pm$0.004 & 176.60$\pm$0.009& 1008.0$\pm$0.004\\
          & 389.05& 2058.0& 1665.1& 1153.1& 966.36& 1719.5& 1840.7& 9624.7\\ 
$[$\ion{N}{ii}$]\, \lambda6584$ & 5.756$\pm$0.150& 56.050$\pm$0.014& 61.840$\pm$0.010& 39.060$\pm$0.011&
36.230$\pm$0.012& 79.590$\pm$0.010& 64.650$\pm$0.025& 348.9$\pm$0.010\\
          &49.257 & 632.30& 590.95& 385.79& 341.23& 641.16&667.69 & 3302.0\\
\ion{He}{i} $\lambda6678$ & 2.133$\pm$0.444& ...& 1.807$\pm$0.437&1.132$\pm$0.264 &0.932$\pm$0.244 &
2.990$\pm$0.256& 5.500$\pm$0.433& 16.000$\pm$0.289\\
          & 17.565& ...& 16.584& 10.732& 8.436& 23.204& 54.478& 145.46\\
$[$\ion{S}{ii}$]\, \lambda6716$ & 3.013$\pm$0.326& 30.670$\pm$0.025&29.380$\pm$0.020 &17.910$\pm$0.021 &
17.340$\pm$0.021& 40.990$\pm$.016& 33.740$\pm$0.044& 174.60$\pm$0.021\\
          & 24.428& 325.52& 265.25& 166.99& 154.35& 313.32&328.57 & 1561.6\\
$[$\ion{S}{ii}$]\, \lambda6731$ & 2.532$\pm$0.388&24.600$\pm$0.031 &22.710$\pm$0.026 &14.910$\pm$0.026 &
13.960$\pm$0.026& 33.650$\pm$0.019& 27.240$\pm$0.054& 140.40$\pm$0.026\\
          & 20.405& 259.33& 203.74& 138.13& 123.49&255.71 & 263.54& 1247.8\\
E($B-V$) (mag)& 0.923 & 1.042& 0.970& 0.490&0.964 & 0.897& 1.004& 0.966\\
N$_{\mathrm e}$ (10$^2$ cm$^{-3}$) & 2.4 & 1.7 & 1.2 & 2.3 & 1.8 & 2.1 & 1.8 & 1.8 \\
            \noalign{\smallskip}
            \hline
         \end{tabular}
     $$ 
   \begin{list}{}{}
   \item[$^{\mathrm{a}}$] Fluxes are in units of 10$^{-16}$ ergs s$^{-1}$ cm$^{-2}$ \AA$^{-1}$ and are listed
   along with (dimensionless) relative errors.
   \end{list}
    \end{table*}

 \begin{table}
 \centering
      \caption[]{Diagnostic emission-line ratios}
         \label{ratios}
 $$
   \begin{tabular}{llll}
            \hline
	    \hline
            \noalign{\smallskip}
 Object Id. & $\log$($[$\ion{O}{iii}$]$/H$\beta$) & $\log$($[$\ion{N}{ii}$]$/H$\alpha$) &
 $\log$($[$\ion{S}{ii}$]$/H$\alpha$) \\
	    \noalign{\smallskip}
            \hline
            \noalign{\smallskip}
 Part 1& $+$0.02 & $-$0.89 & $-$0.94 \\
 Part 2& $-$0.50 & $-$0.51 & $-$0.55\\
 Part 3& $-$0.66 & $-$0.45 & $-$0.55 \\
 Part 4& $-$0.50 & $-$0.48 & $-$0.58 \\
 Part 5& $-$0.67 & $-$0.45 & $-$0.54 \\
 Part 6& $-$0.77 & $-$0.43 & $-$0.48\\
 Part 7&  ... & $-$0.44 & $-$0.49 \\
 Total & $-$0.58 & $-$0.47 & $-$0.54\\
            \noalign{\smallskip}
            \hline
         \end{tabular}
     $$ 
   \end{table}

\subsection{Spectroscopy: Radial velocity determination}

The inspection of the spectrum before the background subtraction, allowed us the detection of
the H$\alpha$ emission-line (and of most other emission lines at lower signal level) all along the slit. 
Since the extended nebula occupies the whole field-of-view,
we conclude that this emission, well outside the visible boundaries of the \ion{H}{ii} region described 
above, stems from the extended, low surface brightness nebula filling the field of view.
Therefore, we measured the position of the H$\alpha$ line in 10 pixel bins 
along the whole slit, in order to establish whether the bright \ion{H}{ii} region and the surrounding 
nebula of excited gas are only 
projected onto each other or have comparable radial velocities, i.e. are spatially coincident.  
Radial velocity measurements were based on the H$\alpha$ line, since this is by far more intense than 
all other detected emission lines, and thus gives a better accuracy.
To estimate the error in radial velocity measurements and detect/correct for possible systematic errors, we
measured the [\ion{O}{i}] sky-lines in the same apertures all along the slit.
We actually found a systematic error in radial velocity with a well defined, regular trend along the slit,
being nearly zero in the central part of the frame (where the spectrum of the bright \ion{H}{ii} region is
located) and increasing when moving toward the borders. The trend
was confirmed also by the measurement of other sky-lines within the wavelength range. Radial velocities
measured in each 10-pixel bin were corrected for the relevant systematic error\footnote{The spectral
resolution does not allow us to separate the geocoronal H$\alpha$ line. However, since this line is at rest 
with respect to Earth like all other night-sky lines, the correction involving the use of the [\ion{O}{i}] 
sky-lines implicitly takes into account the geocoronal line as a radial velocity zero-point. As an additional
remark we note that the emission-line ratios typical of \ion{H}{ii} regions measured in Section 2.4 suggest 
that the contribution of the geocoronal line to the observed H$\alpha$ emission is marginal in this spectrum.}.
The dispersion of position measurements of the brightest sky-lines around their expected values, within
each 10-pixel bin, gives an estimate of the uncertainty of the radial velocities. This is about 12 km s$^{-1}$.
The systemic velocity of the \ion{H}{ii} region, obtained by averaging the radial velocities measured 
within its borders (as visible in the R-band image and in the H$\alpha$ profile), is V$_{\mathrm {LSR}}$ = $-$71
km s$^{-1}$. The extended part of the nebula exhibits a comparable value (within measurement errors) of
radial velocity, on average V$_{\mathrm {LSR}}$ = $-$60 km s$^{-1}$.We conclude that the two structures 
are physically associated and not purely overlapping because of projection effects.  
In principle, part of the detected
H$\alpha$ line might stem from other intervening Galactic nebulae along the line-of-sight or from diffuse
Galactic H$\alpha$ emission. In fact faint diffuse Galactic H$\alpha$ emission appears to cover almost 
the entire sky \citep{r90}. The resolution of our spectra would not allow to unambiguously separate a 
different velocity, weaker H$\alpha$ component, which would appear in narrow blend with the observed line.
However, we do not observe any hint of a second component in the line profile (e.g. double peaks, wings,
humps, etc.), therefore we argue that the emission line component at the observed 
V$_{\mathrm {LSR}}$ = $-$60 km s$^{-1}$ is (if not the only) the dominant one along the line of sight. 
Additionally, as we show below, the spectrum of the extended nebula is typical of classic \ion{H}{ii}
regions, therefore significantly different from the spectrum of diffuse warm ionized
interstellar gas \citep{r90}.

Following the method outlined by \citet{bb93} and using their best fit of the Galactic rotation curve
and our measurement of V$_{\mathrm {LSR}}$, we estimated the distance of the photoionized nebula. 
We assumed the position of the bright 
clump as representative of the center of the nebula, and used the radius of the clump ($\sim$ 16\arcsec) 
as an indication of the positional error.
We used for the Sun the IAU-recommended values of Galactocentric distance and circular velocity, i.e.
R$_0$ = 8.5 kpc and $\Theta_0$ = 220 km s$^{-1}$.
We found a heliocentric distance $d$ = 8.2$^{+2.3}_{-1.8}$ kpc. Since the semi-major and semi-minor axes 
of the nebula, as visible on the DSS2 image, are 15\farcm8 and 14\farcm4, this translates into an 
approximate diameter of 72 pc.
At $d$ $\approx$ 8.2 kpc and b $\approx$ $+$1\fdg83\ this \ion{H}{ii} region is located at 
$z$ $\approx$ 260 pc.

\subsection{Spectroscopy: Physical properties of the extended nebula}

\begin{figure}
\centering
\includegraphics[width=\linewidth]{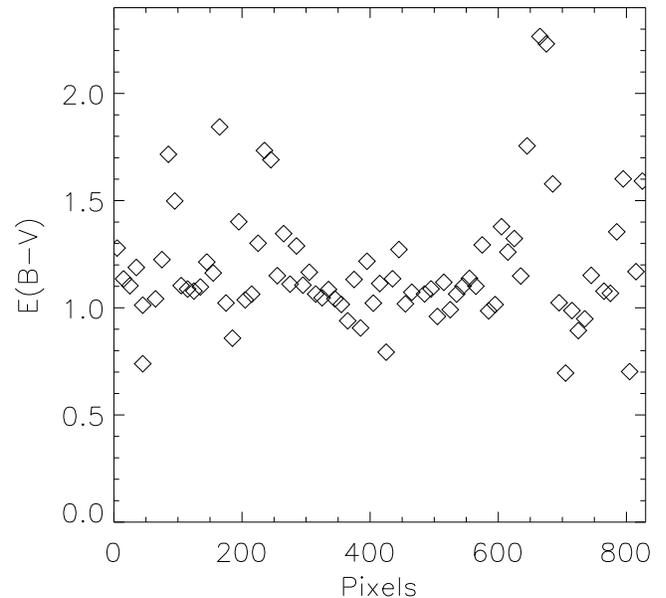} 
\caption{Total foreground extinction obtained from the H$\alpha$/H$\beta$ ratio measured in 10-pixel bins
along the slit, assuming a case B recombination nebula.}  \label{ebv}
\end{figure}

In order to understand the nature of the extended nebula, we also analyzed the trend of the emission-line
ratios along the slit. We found no substantial variation of the diagnostic emission-line ratios, i.e. 
the low surface brightness extended nebula as well shows typical characteristics of classic 
\ion{H}{ii} regions.
The only difference between the fainter and the brighter parts of the nebula falling in the slit resides 
in the ratio of the [\ion{S}{ii}] $\lambda$6716,6731 lines, which in the fainter parts increases to values 
$\ga$ 2, indicating that the extended nebula is in the low density limit, with N$_{\mathrm e}$ $<$ 10
cm$^{-3}$. The total foreground extinction measured all along the slit, as shown in Fig.~\ref{ebv}, does not 
differ substantially from that derived above for the bright clump, ranging from $\sim$ 0.5 to $\sim$ 2.3 mag 
with an average value $<$E(B$-$V)$>$ = 1.2$\pm$0.3 mag. The H$\alpha$ emission measure calculated for the region 
of the bright clump falling within
the slit, after reddening correction, is EM $\sim$ 6.94 $\times$ 10$^3$ pc cm$^{-6}$, while the value 
found for the adjacent parts of the extended nebula is EM $\sim$ 5.33 $\times$ 10$^3$ pc cm$^{-6}$.
We exclude the possibility that the low surface brightness nebula is a region of diffuse ionized gas (DIG)
both on the basis of its H$\alpha$ EM $>>$ 100 pc cm$^{-6}$ and of its [\ion{S}{ii}]/H$\alpha$ ratio, 
which is usually found to be considerably higher in DIG than in classic \ion{H}{ii} regions 
\citep[see, e.g.][]{gwb99}.

Assuming on average an EM $\approx$ 5.3 $\times$ 10$^3$ pc cm$^{-6}$ in the brighter area of the extended
nebula ($\sim$ 334 arcmin$^2$, as estimated on the DSS2 image), considering negligible the contribution of the fainter parts of the 
nebula, and using the distance determined above, we obtained a rough estimate of the total (extinction-corrected)
H$\alpha$ luminosity, L(H$\alpha$) $\sim$ 1.1$\times$10$^{38}$ erg s$^{-1}$. This luminosity is equivalent
to a total number of ionizing photons Q$_{\mathrm {ion}} \, \sim$ 7.7$\times$10$^{49}$ photons s$^{-1}$ \citep{ken98}.
The estimated Q$_{\mathrm {ion}}$ could be produced by a single ionizing star of spectral type
between O5.5Ia and O3V, according to values tabulated by \citet[][ their Tables 5 to 7]{v96}. However,
the corresponding radius of the Str\"omgren sphere would be $\sim$ 6.2 pc, much smaller than the observed
radius ($\sim$ 36 pc) of the ionized nebula. We argue that, as an alternative, the necessary amount of 
ionizing photons
could be emitted by an association of lower luminosity OB stars. As an example, about 50 stars of spectral
types between O9V and B0.5V would produce a total amount of ionizing photons similar to our estimated
values. This interpretation could account for the relatively large size of the Str\"omgren sphere.
Additionally, we note that the Str\"omgren sphere radii given by \citet{v96} are calculated assuming
a hydrogen density N$_{\mathrm H}$ = 10$^2$ cm$^{-3}$ and therefore are significantly smaller than those
tabulated by \citet{ost89} under the assumption of N$_{\mathrm H}$ = N$_{\mathrm e}$ = 1 cm$^{-3}$.
Our spectral analysis showed that, while the bright 1\arcmin-size clump has N$_{\mathrm e}$ $\sim$
2$\times$10$^2$ cm$^{-3}$, the surrounding nebula is in the low-density limit, N$_{\mathrm e}$ $<$ 10
cm$^{-3}$. This fact might contribute to explain the discrepancy between the observed size of the 
ionized region and the tabulated values of Str\"omgren sphere radii. Our conclusion is that the 
nebula can probably be ionized by a few to a few tens of OB stars.

\section{Comparison with FIR and radio emission}

\begin{figure}
\centering
\includegraphics[width=\linewidth]{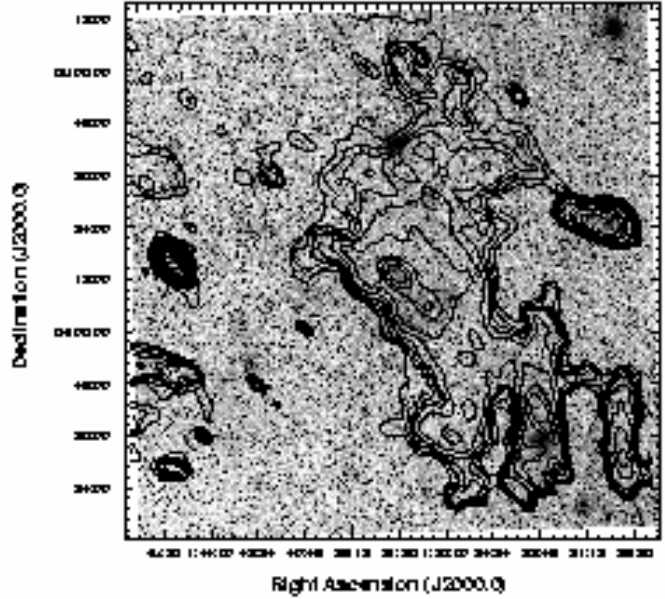} 
\caption{DSS2 image of a 2\degr$\times$2\degr\ field centered on the \ion{H}{ii} region
studied in this work, with overlapping contours of the IRAS 60 $\mu$m image. Contour
levels 1, 2, 3, 4, 6, 10, 20, 30, 50, 100, 200$\sigma$ above the background are drawn.
The $\sigma$ value measured on portions of the image as free as possible from
sources is 0.32 MJy per steradian.}  \label{opt2deg+fir}
\end{figure}

\begin{figure}
\centering
\includegraphics[width=\linewidth]{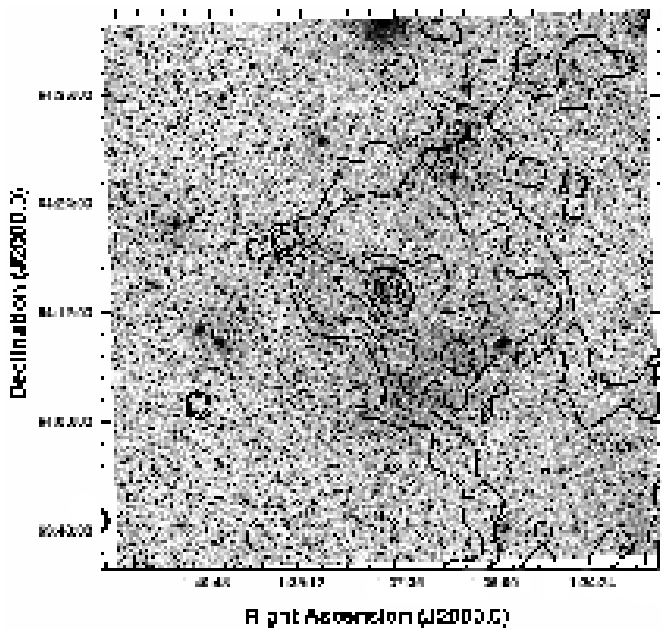} 
\caption{DSS2 image of a 1\degr$\times$1\degr\ field centered on the \ion{H}{ii} region
studied in this work, with overlapping contours of the high resolution IRAS 60 $\mu$m image. Contour
levels 3, 4, 6, 10, 20, 30, and 50 $\sigma$ above the background are drawn.
Measured $\sigma$ is 1.2 MJy per steradian.}  \label{opt+MEfir}
\end{figure}

\begin{figure}
\centering
\includegraphics[width=\linewidth]{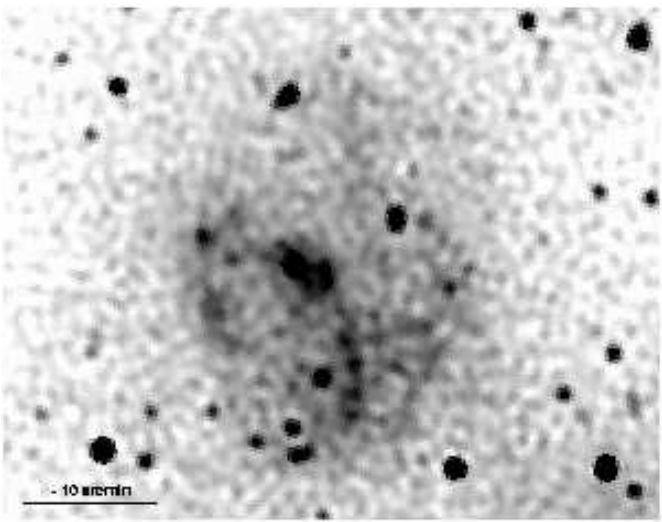} 
\caption{CGPS radio continuum map at 1420 MHz, centered at the position of the optical nebula.
North is up, East to the left. Note the striking similarity between the substructures appearing in this image
and those in the DSS image in Fig.~\ref{dssR}.}  \label{radio_cont}
\end{figure}

In order to derive additional information on the observed nebula, we compared the optical images (both
DSS2 images and our own R-band exposure) with FIR emission from IRAS data and radio emission available
from several sources in the literature and from archival data of the CGPS in the same region of the sky.
Contour maps of the 60 $\mu$m emission, reconstructed from IRAS raw data available at the Groningen
IRAS server, after correction for zodiacal emission, are shown in Fig.~\ref{opt2deg+fir}, overplotted 
onto the
DSS2 R-band image of a 2\degr$\times$2\degr\ field centered at the position of the \ion{H}{ii} region.
As already noticed by other authors \citep{ft96}, besides the two sources listed in the IRAS Point 
Source Catalogue,
extended emission is clearly visible. A striking coincidence of such an extended emission with the 
extended nebula visible in the DSS image strongly suggests that the two features are correlated.
This spatial coincidence is even better visible in Fig.~\ref{opt+MEfir}, where a 1\degr$\times$1\degr\ 
DSS image centered on the optical source is overlapped with the contours of a 60 $\mu$m high
resolution image obtained by processing the IRAS data with a program based on the maximum entropy 
algorithm MeMSys5 \citep{bkk94}. In particular the two strong sources \object{IRAS 01342+6358}  and 
\object{IRAS 01330+6351} are 
coincident with the two main subfeatures visible in the optical nebula. The bright \ion{H}{ii} region 
we identified within the extended nebula lies in close vicinity of the peak of IRAS emission at the center 
of the field.
The spatial coincidence of the extended FIR emission with the extended radio continuum emission detected
by the CGPS has already been shown by \citet{ci03}. Additionally, we notice a remarkable similarity
between the radio and optical substructures of the extended nebula, as can be seen by comparing
the CGPS radio map at 1420 MHz (Fig.~\ref{radio_cont}) with the DSS image of the nebula 
(Fig.~\ref{dssR}, left).
Finally, the radio source \object{NVSSJ013746+641452}, $\sim$ 80\arcsec\ in extent, with integrated flux density of 
20.7$\pm$2.5 mJy at 1.4 GHz, detected by \citet{co98}, coincides, within the positional error, with 
the bright optical clump, which shows a comparable extent in the R-band image.
In view of these spatial coincidences, we suggest that all the above features in the optical, radio, 
and FIR
are related one another. Given the radial velocity measured for the \ion{H}{ii} region and surrounding 
ionized nebula, the above considerations would pose all the above structures at Galactocentric distance 
$\approx$ 15 kpc (i.e. heliocentric distance 8.2 kpc), in contrast with the considerably smaller distance 
proposed by \citet{ci03}, 
based on structures detected in the \ion{H}{i} 21 cm line at V$_{\mathrm {LSR}}$ $\sim$ $-$29.5 km s$^{-1}$.

\begin{figure}
\centering
\includegraphics[width=\linewidth]{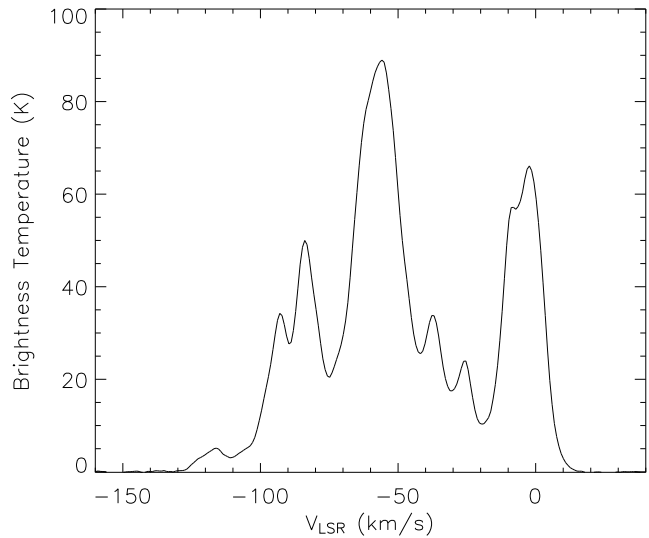} 
\caption{Average \ion{H}{i} emission spectrum within a $\sim$ 30\arcmin$\times$30\arcmin\ box centered at the
position of the optically bright clump.}  \label{HI}
\end{figure}

\begin{figure}
\centering
\includegraphics[width=\linewidth]{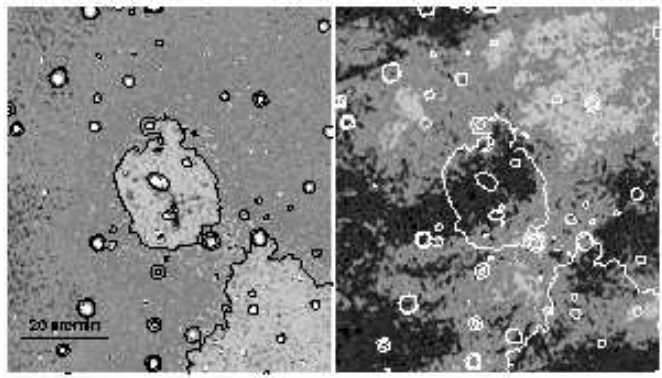} 
\caption{\emph{Left:} Radio continuum image at 1420 MHz, centered at the position of the ionized nebula, overlaid
with a contour indicating the extent of the nebula and its peak of brightness. \emph{Right:} \ion{H}{i} 21 cm
map of the same region, averaged over the velocity range $-$65.8 to $-$69.1 km s$^{-1}$ and overlaid with the
contour of the radio-continuum emission. Darker gray levels indicate lower brightness.}  \label{radiomap}
\end{figure}

The examination of the 21 cm data from the CGPS archive gives further support to our interpretation.
In fact, the plot of brightness temperature in the direction of the \ion{H}{ii} region (Fig.~\ref{HI})
shows a secondary depression in the \ion{H}{i} distribution at V$_{\mathrm {LSR}}$ $\sim$ $-$70 km s$^{-1}$,
in good agreement with our measured radial velocity.
The \ion{H}{i} 21 cm-line image at the position of the extended
nebula shows a hole in the \ion{H}{i} distribution in clear correspondence of the ionized gas visible in the
radio-continuum and optical images. The size and boundaries of the ``hole'' are in good agreement with those of
the ionized nebula, with a ``ring'' of neutral hydrogen largely surrounding the photoionized gas. 
This correspondence is well visible in several velocity channels in the range $\sim$ $-$64 to
$-$71 km s$^{-1}$ (especially at $\sim$ $-$67 km s$^{-1}$) and is shown in Fig.~\ref{radiomap}, 
where the radio-continuum at the position of the nebula is compared to the 21 cm-line map, averaged over 
the velocity interval $-$65.8 to $-$69.1 km s$^{-1}$. These velocities are in good agreement (within
measurement errors) with the radial velocities obtained from the optical spectrum of the nebula.

\section{Conclusions}

We have performed $R$-band photometry and optical spectroscopy of an emission nebula located $\sim$ 2\degr\
above the Galactic plane with the main aims to examine the physical conditions of its gas and establish
its location within the Galactic structure. It was also our intention to establish whether the
observed optical structures in this region of the sky are physically connected one to the others or are
rather distinct structures seen in projection along the line-of-sight.

Our data indicate that the bright condensation $\sim$ 1\arcmin\ in size has typical emission-line ratios
of a classic \ion{H}{ii} region. The radial velocity of the observed H$\alpha$ line suggests that the
\ion{H}{ii} region is located far out in the Galaxy, well beyond the Perseus arm. We speculate that it might
belong to the outer arm of the Galaxy. 
Actually, indications of the presence of spiral arm tracers in the second Galactic quadrant at distances 
up to 11 kpc have been reported by several authors \citep[see, e.g.][ and references therein]{kw89}.
In particular, \citet{kw89}, collecting a number of different spiral arm tracers, gave evidence of the
existence of an outer spiral arm, bending above the Galactic plane in the second quadrant
\citep[see also the structure of the Galactic warp shown e.g. by ][]{w01}, at distances 
comparable to the kinematic distance of the \ion{H}{ii} region discussed in this work. 

The presence of a depression in the \ion{H}{i} profile along the
line-of-sight to the \ion{H}{ii} region at velocities in agreement with those measured from our optical
spectra further supports our interpretation. Additionally, we observe fainter emission lines also
from the extended optical nebula that is completely filling our field-of-view. The radial velocity of
these emission-lines is in agreement  with that of the identified \ion{H}{ii} region, within the 
measurement errors. Once again the diagnostic emission-line ratios are typical of classic \ion{H}{ii} regions.
These facts support the idea that the bright clump and the irregularly shaped,
extended optical nebula are physically related and constitute a center-brightened \ion{H}{ii} region.
Considerations on the emission measure, apparent size, and estimated distance of the nebula lead
to conclude that its gas is ionized by an association of a small number of OB stars.
A comparison between the DSS2 image of this extended nebula and the 60 and 100 $\mu$m IRAS images
reveals a striking spatial coincidence of the structures and, even more strikingly, its optical 
substructures are well reproduced by substructures visible in the 1420 MHz continuum. 
This is in contrast with the recent work
of \citet{ci03}, who give some evidence of the association of these FIR sources and their radio-continuum
counterparts with a nearby (distance $\sim$ 3 kpc) region photoionized by the O star \object{HD 10125}.
Although the arguments given by \citet{ci03} look convincing, they are not supported by direct 
radial velocity measurements of the ionized gas. Indeed, the correspondence we find between a ``hole'' in
the \ion{H}{i} distribution at velocities around $-$67 km s$^{-1}$ and the extended radio-continuum emission
appears much tighter than the correspondence with an \ion{H}{i} ``hole'' at $-$29.5 km s$^{-1}$ suggested by
\citet{ci03}.
In the light of our observations, we argue that both 
the FIR sources and the radio-continuum emission are rather associated to the more distant ($d \sim$ 8.2
kpc) photoionized nebula we have identified. 

\begin{acknowledgements}
ST is grateful to S. Ciroi for helpful discussions. We thank the referee, D. Russeil, for his timely and 
useful report. A part of this work was supported by the Austrian
Science Fund (FWF) under project P15065.
This work made use of data produced by the CGPS Consortium. The Canadian Galactic Plane Survey (CGPS) 
is a Canadian project with international partners. The Dominion Radio Astrophysical Observatory is 
operated as a national facility by the National Research Council of Canada. The CGPS is supported by 
a grant from the Natural Sciences and Engineering Research Council of Canada.
IRAS data were obtained from the IRAS Data Server located at Kapteyn and operated by SRON.

\end{acknowledgements}

\end{document}